\documentstyle[12pt]{article}
\begin{document}
\parskip          =0mm
\oddsidemargin    =1cm
\evensidemargin   =0cm
\textwidth        =15cm
\textheight       =20.5cm
\headheight       =0cm
\topskip          =0cm
\topmargin        =0.5cm


\title{SOME POINTS ON CASIMIR FORCES}
\author{A.H. Abbassi$^1$ $\;\&\;$ A.R. Sadre Momtaz$^2$\\
{\small $^1$ Department of Physics, School of Sciences, Tarbiat Modarres University,} \\
{\small P.O.Box 14155-4838, Tehran, Iran.}\\
{\small $^2$ Department of Physics, School of Sciences, Gilan University,
Rasht,Iran.}}
\date{}
\maketitle
\begin{abstract} 
Casmir forces of massive fermionic Dirac fields are calculated for
parallel plates geometry in spatial space with dimension $d$ and 
imposing bag model boundary conditions. It is shown that in the range 
of $ma\gg 1$ where $m$ is mass of fields quanta and $a$ is the separation distance of the plates, it is equal to massive bosonic fields
Casimir force for each degree of freedom. We argue this equality exists
for any massive anyonic field in two-dimensional spatial space. 
Also the ratio of massless fermionic field Casimir force to its
bosonic correspondent in $d$-dimensional spatial space is
$(1-\frac 1{2^d})$.\\

\smallskip
\noindent PACS: 05.30-d, 05.30-Fk, 03.70+k\\

\bigskip
  
\end{abstract}
\begin{center}
{\bf Introduction}
\end{center}

 Casimir effect of vacuum fluctuations has been extensively investigated in various contexts[1-8]. The calculation for massive bosonic field between two parrallel and confining plates has been carried out [9-11]. In the range of $ma\gg 1$ ,the corresponding Casimir force is attractive and is equal to $\frac{m^{d/2+1}}{(4\pi a)^{d/2}}\exp{(-2ma)}$ per degree of freedom, where d stands for the spatial space dimension. The calculation of this effect has also been carried out for zero-mass fermionic fields by Ken Johnsen and others for the first time [12-13]. It turns out that the contribution of each degree of freedom of zero-mass fermionic field in the Casimir force is $\frac 78$ the contribution of the corresponding photonic field. It is well known in statistical physics that the pressure due to free fermion and boson degree of freedom at finite temperature is related by the same $\frac 78$ factor[14]. In the first part of this article ,we prove that in the range of $ma\gg 1$,the contribution of each degree of freedom of massive fermionic field in Casimir force is the same as the corresponding massive bosonic field in any spatial space dimension.
The second part of the discussion is restricted to two dimensional satial spaces and concludes that this equality is conserved by any massive anyonic field which exists only in a two-dimensional spatial space. The third part of the calculation is confined to massless fields in $d$-dimensional spatial spaces and the ratios of these forces are calculated in this case.
\begin{center}
{\bf Massive Fermionic Fields}
\end{center}

The Hamiltonian for a quantized massive fermionic field between two parrallel plates in $d$-dimensional spatial spaces is 
\begin{equation}
H=(\frac{DRF}4)\sum\limits_{\lambda ,k}\int \frac{d^{d-1}k_T}{(2\pi)
^{d-1}}[b^{+}_{\lambda k k_T} b_{\lambda k k_t}-d_{\lambda k k_T}
d^{+}_{\lambda k k_T}]\omega_{kk_T}
\end{equation}\label{1}                                                                                              
where $\lambda =\pm 1$ is a spin index,$k$ is quantized momentum normal to plates; $k_T$ is a $(d-1)$-dimensional continuous transverse momentum; $\omega _{kk_T} =(k^2 + k_T ^2 +m^2)^{1/2}$ and  $\hbar =1$ is assumed . DRF stands for dimensional regularization factor [15]. Here the oparator $b_{\lambda kk_T}b^+_{\lambda kk_T}$ counts the number of fermions in the mode $(\lambda,k,k_T)$ while $d^+_{\lambda kk_T}d_{\lambda kk_T}$ counts the number of antiparticles in the mentioned mode. Since the $d^+_{\lambda kk_T}d_{\lambda kk_T}$ operator appears in the Hamiltonian, the vacuum state will have a divergent zero-point energy which gives rise to the Casimir effect. On the two cofining plates,the field is required to satify the MIT boundary conditions [16],leading to 
\begin{equation}
k=\frac{n\pi}{2a} \;\;\;\;\; with\; n=1,3,5,...
\end{equation}\label{2}                                                                                              
                                                                                              
The expectation value of $H$ in vacuum state is
\begin{equation}
<0|H|0>=-(\frac{DRF}4)\sum\limits_{\lambda ,k}\int\frac{d^{d-1}k_T}
{(2\pi)^{d-1}}\omega_{kk_T}
\end{equation}\label{3}                                                                                              

In a homogeneous and isotropic $d$-dimensional space we have[9]. 
\begin{equation}
\int f(k)d^dk=\frac{2\pi^{d/2}}{\Gamma(d/2)}\int k^{d-1}f(k)dk
\end{equation}\label{4}                                                                                              
                                                                                       
Since the $(d-1)$-dimensional plates are homogeneous and isotropic,(4) may be used to cast (3) into
 \begin{equation}
<0|H|0>=-\frac{(DRF/4)}{2^{(d-2)}\pi^{(d-1)/2}\Gamma((d-1)/2)}
\sum\limits_{\lambda ,k}\int\limits_0^\infty k^{d-2}_T
\omega_{kk_T}dk_T
\end{equation}\label{5}                                                                                              
                                               
  Replacing the integration parameter $k_T$ by $[(k^2 +m^2)t]^{1/2}$, (5) becomes
\begin{equation}
<0|H|0>=-\frac{(DRF/4)}{2^{(d-1)}\pi^{(d-1)/2}\Gamma ((d-1)/2)}
\sum\limits_{\lambda ,k} (k^2+m^2)^{d/2}\int\limits_0^\infty
t^{(d-3)/2}(1+t)^{1/2}dt
\end{equation}\label{6}                                                                                              
                           
  Now using the integral form of beta function
\begin{eqnarray}
B(1+r,-S-r-1)=\int\limits_0^\infty t^r(1+t)^Sdt \nonumber              
\end{eqnarray}                                                                                              

where 
 \begin{equation}
B(q,p)=\frac{\Gamma (p)\Gamma (q)}{\Gamma (p+q)}
\end{equation}\label{7}                                                                                              
(6) will be reduced to 
\begin{equation}
<0|H|0>=-\frac{(DRF/4)\Gamma(-d/2)}{2^{(d-1)}\pi^{(d-1)/2}
\Gamma (-1/2)}\sum\limits_{\lambda ,k}(k^2+m^2)^{d/2}
\end{equation}\label{8}
\begin{equation}
<0|H|0>=-\frac{(DRF/4)\Gamma(-d/2)}{2^{(d-1)}\pi^{(d-1)/2}
\Gamma (-1/2)}\sum\limits_{n=1}^\infty \left [\frac
{(\pi(2n-1))^2}{4a^2}+m^2\right ]^{d/2}
\end{equation}\label{9}                                                                                              
  Since
 \begin{equation}
\sum\limits_{n=1}^\infty \left [\frac
{(\pi(2n-1))^2}{4a^2}+m^2\right ]^{d/2}=\sum\limits_{n=-\infty}^{n=0} \left [\frac
{(\pi(2n-1))^2}{4a^2}+m^2\right ]^{d/2}
\end{equation}\label{10}                                                                                              
                                                                                           (9) can be arranged in the form
 \begin{equation}
<0|H|0>=-\frac{(DRF/4)\Gamma(-d/2)}{2^{(d-1)}\pi^{(d-1)/2}
\Gamma (-1/2)}\frac 12\sum\limits_{n=-\infty}^{+\infty} \left [\frac
{(\pi(2n-1))^2}{4a^2}+m^2\right ]^{d/2}
\end{equation}\label{11}                                                                                              
 For determining the infinite series in (11) we compute it as follows.
Let us consider the theta function  $\nu_2(x)$ which is defined as[17-18]
\begin{equation}
\nu_2(x)=\sum\limits_{n=-\infty}^{+\infty}\exp{\left [ -\pi(n-
\frac 12)^2x\right ]}
\end{equation}\label{12}                                                                                                                                                                      
and define $S$ as
\begin{equation}
S=\int\limits_0^\infty dx\; x^{-(d+2)/2}\exp{(-\frac{m^2}{\pi}x)}
\nu_2(x/a^2)
\end{equation}\label{13}                                                                                              
  By using the integral form of gamma function, S can be written as
\begin{equation}
S=\pi d/2\Gamma (-\frac d2)\sum\limits_{n=-\infty}^{+\infty}
\left [(\frac{2n-1}{2a})^2+(\frac m\pi)^2 \right ]^{d/2}
\end{equation}\label{14}                                                                                              
  Comparing (14) and (11) gives
 \begin{equation}
<0|H|0>=2^{-d} S(DRF/4)
\end{equation}\label{15}                                                                                              
  Thus calculating $<0|H|0>$ turns to finding $S$. Changing the integration parameter $x$ into $y^{-1}$ in (13) ,$S$ becomes
 \begin{equation}
S=\int\limits_0^\infty dy \; y^{(d-2)/2}\nu_2(\frac 1{a^2y})\exp{(-
\frac{m^2}{\pi y})}
\end{equation}\label{16}                                                                                              
 $\nu_2(x)$ and $\nu_4(x)$ are related by the relation  
\begin{equation}
\nu_2(\frac 1{a^2y})=a\sqrt y \nu_4(a^2y)
\end{equation}\label{17}                                                                                                                                                                                       
where $\nu_4(x)$ is  
\begin{equation}
\nu_4(x)=\sum\limits_{n=-\infty}^{+\infty} (-1)^n\exp{(-\pi n^2x)}
\end{equation}\label{18}                                                                                              
                                                                                          
Now (16) may be rearranged by using (17) and(18) to
 \begin{equation}
S=a\int\limits_0^\infty dy \; y^{(d-1)/2}\exp{(-\frac {m^2}{\pi y})}
[1+\sum\limits_{{n=-\infty}\atop {n\not = 0}}^{+\infty}
(-1)^n\exp{(-\pi n^2a^2y)}]
\end{equation}\label{19}  
\begin{equation}
S=a(\frac{m^2}\pi)^{(d+1)/2}\int\limits_0^\infty dt\; t^{-(d+3)/2}\exp
{(-t)}[1+\sum\limits_{n=-\infty \atop n\not = 0}^{+\infty}
(-1)^n\exp{(\frac{-a^2n^2m^2}t)}]
\end{equation}\label{20}                                                                                              
                                                                                            
  The integral form of gamma function may be used for the first term
 \begin{eqnarray}
S&=&a(\frac{m^2}\pi)^{(d+1)/2}[\Gamma (-\frac{d+1}2)+\nonumber\\
& &\sum\limits_{n=-\infty \atop n\not= 0}^{+\infty}
(-1)^n\int\limits_0^\infty dt\; t^{-(d+3)/2}\exp{[-(t+\frac
{a^2n^2m^2}t)]}]
\end{eqnarray}\label{21}                                                                                              
  Then changing $t\rightarrow|anm|t^\prime$, (21) can be written as
 \begin{eqnarray}
S&=&a(\frac{m^2}\pi)^{(d+1)/2}[\Gamma (-\frac{d+1}2)+ \nonumber\\
& &\sum\limits_{n=-\infty \atop n\not= 0}^{+\infty}
(-1)^n|anm|^{(-d+1)/2}\int\limits_0^\infty dt^\prime\; {t^\prime}^{-(d+3)/2}
\exp{[-|anm|(t^\prime+\frac 1{t^\prime})]}]
\end{eqnarray}\label{22}                                                                                              
 The integral form of modified Bessel functions may be used to simplify (22),we have 
\begin{equation}
K_\nu(x)=\frac 12\int\limits_0^\infty \exp{[-\frac x2(t+\frac 1t)]}
t^{-(\nu+1)}dt
\end{equation}\label{23}                                                                                                                                                                                     and with (23) we get
 \begin{equation}
S=a(\frac{m^2}\pi)^{(d+1)/2}[\Gamma (-\frac{d+1}2)+ 
 \sum\limits_
{n=-\infty \atop n\not= 0}^{+\infty} (-1)^n\frac{K_{(d+1)/2}(2|anm|)}
{|anm|^{(d+1)/2}}]
\end{equation}\label{24}                                                                                              
  By applying (24) to (15) we obtain
 \begin{eqnarray}
<0|H|0>&=&\frac a{2^d}(\frac{m^2}\pi)^{(d+1)/2}[\Gamma (-\frac{d+1}2)+\nonumber\\
& &4\sum\limits_{n=1}^{+\infty} (-1)^n \frac{K_{(d+1)/2}(2|anm|)}
{|anm|^{(d+1)/2}}(\frac{DRF}4)]
\end{eqnarray}\label{25}                                                                                              
  The first term in (25) gives rise to a force independent of "a" and can be dropped.
 \begin{equation}
<0|H|0>=\frac a{2^{d-2}}(\frac{m^2}\pi)^{(d+1)/2}
\sum\limits_{n=1}^{+\infty} (-1)^n \frac{K_{(d+1)/2}(2|anm|)}
{|anm|^{(d+1)/2}}(\frac{DRF}4)
\end{equation}\label{26}                                                                                              
 The asymptotic form of $K_\nu$ is 
\begin{equation}
K_\nu(z)\rightarrow \sqrt{\frac{\pi}{2z}}\exp{(-z)}\;\;\;\; for
z\gg 1
\end{equation}\label{27}                                                                                                                                                                     
In the range  $ma\gg 1$,(27) may be used in (26) and only the first term has a significant contribution and the others can be ignored. Then we have
 \begin{equation}
<0|H|0>=-\frac{(DRF)m^{d/2}}{2^{d+1}a^{d/2}\pi^{d/2}}\exp{(-2ma)}
\end{equation}\label{28}                                                                                              
  By using  $F =-\frac{\partial E}{\partial a}$  and the approximation  $ma\gg 1$, Casimir force can be obtained
 \begin{equation}
F=-\frac{(DRF)m^{\frac d2+1}}{2^d a^{d/2}\pi^{d/2}}\exp{(-2ma)}
\end{equation}\label{29}                                                                                              
 Since we have four degrees of freedom for fermionic fields, for finding Casimir force per each degree of freedom we should divide (29) by (DRF)
\begin{equation}
F_f=-\frac{m^{d/2+1}}{4\pi a^{d/2}}\exp{(-2ma)}\;\;\;\;\;per\; each\; degree
\end{equation}\label{30}                                                                                              
 This force is attractive and has the exact form of Casimir force of massive bosonic fields in the range $ma\gg 1$.
\begin{center}                                                                        {\bf Casimir Force In Two Space}
\end{center}

It is well known that planar physical systems in two space and one time dimensions display a peculiar phenomena. There exist quantum states that cary angular momentum which are quantized in half-integer units and whose statistics are neither bosonic nor fermionic.The excitations with fractional statistics are often referred to as anyons[19].Any non-relativistic system of anyons may be represented by a system of bosons or fermions plus an interaction which is characterized by a Chern-Simon potential. Since the corresponding Lagrangian of Chern-Simon fields is metric independent,its contribution in energy-stress tensor $T^{\mu\nu}$ vanishes. Any quantized massive field may be considered as an equivalent system of non-relativistic gas for which the Chern-Simon theory may be used to described its physical behaviour. Due to the lack of energy-stress tensor in Chern-Simon potentials,we can conclude that it does not contribute in any way to the Casimir force.This can explain why in two dimensions we get the same result for bosonic and fermionic massive fields Casimir forces. Now,as regards the Casimir force of massive  anyonic fields,in the range of $ma\gg 1$ the quantized massive anyonic field may be considered a non-relativistic system of bosons or fermions,plus a Chern-Simon interaction term which has no contribution to the Casimir force. Thus,we can conclude that we arrive at the same result for any massive anyonic field we have obtained for bosonic and fermionic massive fields in the range $ma\gg 1$.
\begin{center}                                                        
{\bf Massless Fermionic Field}
\end{center}
Now,first of all we would like to calculate the Casimir force of massless fermionic field in an arbitrary spatial space dimension and in this way find the ratio of fermionic to bosonic massless Casimir force in general case,which is $\frac 78$ in three dimensions. Using(19) and putting $m=0$ we get         
\begin{equation}
S=a\int\limits_0^\infty dy\; y^{(d-1)/2}[1+\sum\limits_
{n=-\infty \atop n\not= 0}^{+\infty} (-1)^n\exp{(-\pi n^2a^2y)}]
\end{equation}\label{31}                                                                                              
  Again the first term may be dropped because it gives rise to a force independent of a. Changing the integration parameter to $t=\pi n^2a^2y$, (31) gives
 \begin{equation}
S=a\sum\limits_{n=-\infty \atop n\not= 0}^{+\infty}
\frac{(-1)^n}{(\pi n^2a^2)^{d+1)/2}}\int\limits_0^\infty
dt\; t^{(d-1)/2}\exp{(-t)}
\end{equation}\label{32}                                                                                              
 By using the integral form of gamma function,(32)may be rearranged to
 \begin{equation}
S=\frac{2a\Gamma (\frac{d+1}2)}{(\pi a^2)^{(d+1)/2}}
\sum\limits_{n=1} \frac{(-1)^n}{n^{d+1}}
\end{equation}\label{33}                                                                                              
  The infinite series in (31) can be replaced by the following closed form,let
 \begin{eqnarray}
 S^\prime&=&\sum\limits_{n=1}^\infty \frac{(-1)^n}{n^{d+1}}=
-\sum\limits_{n \atop odd} \frac 1{n^{d+1}}+\sum\limits_{n \atop even}
\frac 1{n^{d+1}}\nonumber\\
S^\prime&=&-\sum\limits_{n \atop odd} \frac 1{n^{d+1}}
+\frac 1{2^{d+1}}\sum\limits_{n=1} \frac 1{n^{d+1}}                                                                                             
\end{eqnarray}\label{34}                                                                                              
  Now we use the zeta function which gives
 \begin{equation}
\sum\frac 1{n^{d+1}}=\zeta (d+1)
\end{equation}\label{35}                                                                                              
 and(34) becomes
 \begin{equation}
S^\prime=-\sum\limits_{n \atop odd}\frac 1{n^{d+1}}+\frac{\zeta (d+1)}
{2^{d+1}}
\end{equation}\label{36}                                                                                              
On the other hand $\zeta (d+1)$may be rearranged as
 \begin{equation}
\zeta (d+1)=\sum\limits_{n \atop odd}\frac 1{n^{d+1}}
+\sum\limits_{n \atop even}\frac 1{n^{d+1}}=\sum\limits_{n \atop odd}
\frac 1{n^{d+1}}+\frac {\zeta (d+1)}{2^{d+1}}
\end{equation}\label{37}                                                                                              
  So(37) gives
 \begin{equation}
\sum\limits_{n \atop odd}\frac 1{n^{d+1}}=
(1-\frac 1{2^{d+1}})\zeta (d+1)
\end{equation}\label{38}                                                                                                                                                                                                   $S^\prime$ can  be found by putting (38)in (36),we obtain
 \begin{equation}
S^\prime=(1-\frac 1{2^d})\zeta (d+1)
\end{equation}\label{39}                                                                                              
$S$ can be calculated from (33) by using(39) for $S$
 \begin{equation}
S=-\frac{2a\Gamma (\frac{d+1}2)}{(\pi a^2)^{(d+1)/2}}
(1-\frac 1{2^d})\zeta (d+1)
\end{equation}\label{40}                                                                                              
 (40) and (15) give $<0|H|0>$ as 
 \begin{equation}
<0|H|0>=-\frac{a\Gamma (\frac{d+1}2)}{(\pi a^2)^{(d+1)/2}}
(1-\frac 1{2^d})\zeta (d+1)(\frac{DRF}4)
\end{equation}\label{41}                                                                                              
  Casimir force may be obtained by $F=-\frac{\partial E}{\partial a}$
 \begin{equation}
F=-\frac{d\Gamma (\frac{d+1}2)}{2^{d-1}(\pi a^2)^{(d+1)/2}}
(1-\frac 1{2^d})\zeta (d+1)(DRF)
\end{equation}\label{42}                                                                                              
  For finding Casimir force per degree of freedom we should divide(42) by (DRF) and we get
 \begin{equation}
F_{f-massless}=-\frac{d\Gamma (\frac{d+1}2)}{(4\pi a^2)^{(d+1)/2}}
(1-\frac 1{2^d})\zeta (d+1)
\end{equation}\label{43}                                                                                              
  Casimir force of massless bosonic field has been calculated [6]. It is equal to
 \begin{equation}
F_{b-massless}=-\frac{d\Gamma (\frac{d+1}2)}{(4\pi a^2)^{(d+1)/2}}
\zeta (d+1)
\end{equation}\label{44}                                                                                              
 Dividing (43) by(44) we get the final result,the ratio of fermionic to bosonic massless Casimir force in an arbitrary spatial space with dimension  $d$
 \begin{equation}
\frac{F_{f-massless}}{F_{b-massless}}=(1-\frac 1{2^d})
\end{equation}\label{45}                                                                                              
 In three dimensions,we get the same $\frac 78$ factor as mentioned before and in two dimensions this ratio is $\frac 34$. 
\begin{center}                                                                               {\bf Conclusion}
\end{center}
We have shown that the Casimir forces of massive fields have the same form and the ratio of massless fermionic to bosonic field Casimir force is $(1-\frac 1{2^d})$. This number now needs only a physical explanation.We think this work will help us to find a proper relativistic theory of anyones. It is ineresting to note that this ratio depends on d and is close to one for large  d . This is surpring as one would expect it to be a pure spin dependent property and independent of spatial space parameters.                                                                        
    

\begin{thebibliography}{1}
\bibitem[1]{} Casimir,H.B.G.Proc.Ned.Akad,Wet,B51,793,(1948).
\bibitem[2]{} Brown,L.S. and Maclay,G.J.Phys.Rev.184,1272,(1964).
\bibitem[3]{} Lutken,C.A.and Ravandal,F.Ibid,A31,2082,(1985).
\bibitem[4]{} Sparnaay,M.J.Physica,24,751,(1958).
\bibitem[5]{} Mehra,J.Ibid.,37,145,(1967).
\bibitem[6]{} Pluien,G.et al.Physics Reports,134,87-193,(1986).
\bibitem[7]{} Lamoreaux,S.K.Phys.Rev.Lett.,78,5,(1997).
\bibitem[8]{} Milton,K.A.Phys.Rev.D,55,(8),4940-4946,(1997).
\bibitem[9]{} Ambjorn,J.and Wolferam,S. Annals of Physics,147,1-32,(1983).
\bibitem[10]{} Barton,G.and Dombey,N.Ibid.,162,231-272,(1985).
\bibitem[11]{} Bordag,M.et al.Phys.Rev.D,56,4896-4946(1997).
\bibitem[12]{} Gundersen,S.A. and Ravandal,F.Annals of Physics,182,90-111,(1988).
\bibitem[13]{} Johnson,K.Acta Physica Polonica,B6,865,(1975).
\bibitem[14]{} Hung,K. Statistical mechanics,(2nd edn),p.189.Wiley,(1987).
\bibitem[15]{} Muta,T.Foundations of quatum chromodynamics.Word Sci., 110 ,118-119,(1987).
\bibitem[16]{} Chodos,A. and Thorn,C.B.Phys.Lett.,B53,359,(1974).
\bibitem[17]{} Magnus, W. et al. Formulas and theorems for special functions of mathematical physics,(3rd edn).Springer-Verlag,(1966).
\bibitem[18]{} Elizalde,E. Ten physical applications of spectral zeta function.Springer-Verlag,(1995).
\bibitem[19]{} Alberto,L.Anyons.Springer-Verlag,(1992).
\end{thebibliography}
\end{document}